\documentclass[a4paper,11pt]{article}
\usepackage{amsmath,amsfonts,amssymb,times,mathptmx,graphicx,amsthm,lineno}
\usepackage{placeins}
\usepackage[T1]{fontenc}
\usepackage[latin1]{inputenc}
\title{\bf Positioning Error Probabilities for Some Forms of
Center-of-Gravity Algorithm Calculated with the \\
Cumulative Distributions. Part II
}
\author{Gregorio Landi$^a$\thanks{Corresponding
author. Gregorio.Landi@fi.infn.it}~,   Giovanni E. Landi$^b$\\
\\
\llap{$^a$} Dipartimento di Fisica e Astronomia,
Università di Firenze and INFN\\
Largo E. Fermi 2 (Arcetri) 50125, Firenze, Italy\\
\\
\llap{$^b$} ArchonVR S.a.g.l.,\\
Via Cisieri 3,
6900 Lugano, Switzerland.}
\date{February 25, 2021 }
\begin{document}
\vspace{-45mm}
\maketitle 
\begin{abstract}
To complete a previous work, the probability density functions
for the errors in
the center-of-gravity as positioning algorithm
are derived with the usual methods of
the cumulative distribution functions. These methods
introduce substantial complications compared
to the approaches used in a previous publication
on similar problems.
The combinations of random
variables considered are:
$X_{g3}=\theta(x_2-x_1) (x_1-x_3)/(x_1+x_2+x_3) +
\theta(x_1-x_2)(x_1+2x_4)/(x_1+x_2+x_4)$
and $X_{g4}=(\theta(x_4-x_5)(2x_4+x_1-x_3)/(x_1+x_2+x_3+x_4)+
\theta(x_5-x_4)(x_1-x_3-2x_5)/(x_1+x_2+x_3+x_5)$
The complete and partial forms of the probability density functions of
these expressions of the center-of-gravity algorithms
are calculated for general probability density functions of
the observation noise. The cumulative probability
distributions are the essential steps in this study, never
calculated elsewhere.
\end{abstract}
%
%
%
\tableofcontents 
\pagenumbering{arabic} \oddsidemargin 0cm  \evensidemargin 0cm

\section{Introduction}

Specialized probability density functions (PDFs) are
essential instruments for track fitting in heteroscedastic
systems, that must be constructed for the
positioning algorithms of the track
observations (hits). The demonstrations of
refs~\cite{landi08,landi09} require the appropriate variances for
each hit of the detector to obtain optimal fits.
The variance calculations and the search of the
maximum likelihood are essentially based on
these PDFs, and, being non-Gaussian, the results of the
maximum likelihood add further improvements to
the track resolution beyond the least-squares method.
In fact, the PDFs of the most frequently used positioning algorithms
turn out very different from Gaussian PDFs, with
Cauchy-Agnesi-like tails due to the non-linearity of the
algorithms. In principle, the PDFs have infinite variances that is
improperly handled by methods based on variance minimizations, hence,
drastic modification must be introduced to extract finite variances.
References~\cite{landi10,landi12} was devoted to obtain the error PDFs for
some forms of the center of gravity (COG) positioning algorithms with a the
simplified method of ref.~\cite{landi06}.
That method obtains a PDF without using the cumulative
distribution, essential instead in the standard method~\cite{gnedenko}.
To complete that result, this paper is devoted to compute
the PDFs of the COG algorithms of ref.~\cite{landi12} with the use of
the cumulative distribution, as we did in ref.~\cite{landi11}
for the PDFs the COG algorithms of ref.~\cite{landi10}.
The calculation of the cumulative distribution is very
long and complicated, requiring many integrations
on sectors of the space of the random variables
given by the noisy strip signals.
However, this approach was the first one we used to compute
the PDFs  extensively utilized in ref.~\cite{landi05} and
in all the searches of likelihood maximums.
The following sections are
our summary notes of those developments.
We can not hide the complexity of the likelihood
method and the extraction of the individual
hit error PDFs to insert in the
likelihood. The gain in resolution observed
in the simulations of refs.~\cite{landi06,landi05,landi07}
largely justifies these
increases of complexity. In addition to this,
the demonstrations of refs.~\cite{landi08,landi09}
show the non-optimality of the use of the standard
least squares method (optimum only for
homoscedastic systems) obliging to utilize
more elaborated procedures.

To illustrate the power of these procedures,
we introduced in ref.~\cite{landi07} a very simple toy model
to clearly show the drastic gains in the parameter
resolution compared to the standard least squares.
The toy model requires few lines of
MATLAB-code~\cite{matlab} to be implemented
and could be an easy teste of heteroscedasticity.
In fact, that toy model was extensively explored
in ref.~\cite{rudi} well outside the range of our
parameters with very interesting results.
Instead, ref.~\cite{bernard} sustains that the results of the
toy model are interesting only for a
small number of observations. Going to large numbers
one obtains results {\em compatible with common wisdom. Neither
homoscedasticity nor heteroscedasticity are found to play any
role in the matter.} This statement of~\cite{bernard} is really
surprising, given that, inserting its plots of the two fit types
in the same figure, the direct comparison clearly shows large
differences in rate of growth and amplitude.
%
All in favor of
heteroscedasticity,
%
%
albeit with a rate of growth not as
rapid as that for the observation numbers of ref.~\cite{landi07}.
Evidently, the addition of a good observation around to hundred of them
has a much less effect than around to ten. However, those large
numbers have no physical meaning for our aim of particle tracking.

To handle properly the intrinsic heteroscedasticity of a tracker
system, the following PDFs are only first steps. Other tools
are essential and require further details,
shortly outlined in ref.~\cite{landi06,landi05}.
For evident reason, all these required developments are split in
parts and published separately.

\section{The discontinuity around $x\approx 1/2$ of the three strip COG (COG$_3$)}

Reference~\cite{landi11} (and ref.~\cite{landi10}) account for the
COG$_3$ in its simplest form, without the introduction
of the effects of the noise in promoting a
lateral strip to become the seed strip (that with the maximum signal).
This noise effect produces gaps/discontinuities in the
PDF of the COG$_3$ algorithm, gaps/discontinuities essential
for a consistent likelihood.
The COG$_3$ histograms show, for large
signal distributions, reductions of the data density around
$\pm 1/2$ as discussed in ref.~\cite{landi12} and with more
details in refs.~\cite{landi01,landi03}.

The PDFs of the COG$_3$ around the discontinuity of
$x_{g3}\approx 1/2$ will be calculated with the standard method~\cite{gnedenko}.
As first step we have to obtain the cumulative distribution,
its differentiation gives the PDF.
In this case, the combined probability of four stirps
has to be considered. The three strips of the COG$_3$ in
the range $-1/2\ll x_{g_3}\ll1/2$ and the forth strip
when the $x_{g3}\approx 1/2$ and the noise is can move
the strip with the maximum signal to the strip to its right.
Or to its left for $x_{g3}\approx -1/2$.

To save the conventions of the previous approaches we will
indicate the strips of our interest (from left to right )
$<3>,<2>,<1>,<4>$. The gap/discontinuity is present when
the maximum-strip signal moves from the strip $<2>$ to the
strip $<1>$ and the COG$_3$, formerly calculated with signals
of the strips $<3>,<2>,<1>$, must be
calculated with signals of $<2>,<1>,<4>$.

\subsection{Sign conventions}

The position of the strips are defined relatively to
the strip with the maximum signal, negative positions to the left,
positive positions to the right. When the strip with
maximum signal is $<2>$ and $\{x_i,i=1,2,\ldots\}$ are
the signals of the strips, the COG$_3$ expression $x_{g3}$ is:
\begin{equation}\label{eq:equation_1}
    x_{g3}=\frac{x_1-x_3}{x_1+x_2+x_3}
\end{equation}
When $x_1$ is the maximum strip signal, $x_{g3}$ is given by:
\begin{equation}\label{eq:equation_2}
    x_{g3}=\frac{x_4-x_2}{x_1+x_2+x_4}
\end{equation}
This convention is impractical for our needs because when
$x_2\approx x_1>> x_3,x_4$ the previous definition generates a
transition of $x_{g3}$ from $1/2$ to $-1/2$. It is fundamental to use  in any case
the center of the strip $<2>$ as origin of the COG reference system.
Now eq.~\ref{eq:equation_2} becomes:
\begin{equation}\label{eq:equation_2a}
    x_{g3}=\frac{x_4-x_2}{x_1+x_2+x_4}+1=\frac{x_1+2 x_4}{x_1+x_2+x_4}
\end{equation}
and the transition of the leading strip increases the values of $x_{g3}$ beyond $1/2$.
The (almost always) positive value of $x_3$ or $x_4$ impedes to $x_{g3}$ to be equal to
$1/2$. Negative values of the strip signals allow $x_{g3}=1/2$, evidently the
negative values of $x_i$ are produced by the noise for very low signals.

\begin{figure} [h!]
\begin{center}
\includegraphics[scale=0.5]{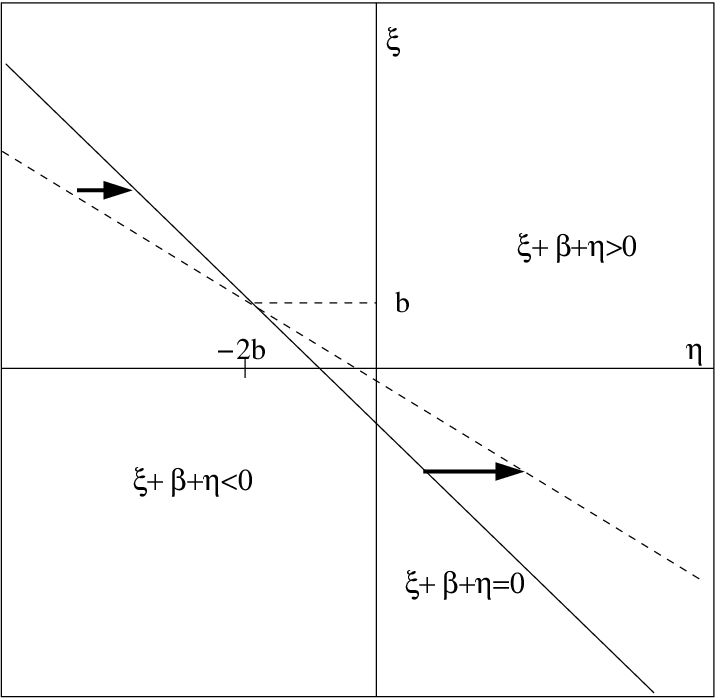}
\caption{\em Sector of the plane $(\eta,\xi)$ where $\xi+\eta+\beta>0$ or
$\xi+\eta+\beta<0$ and the boundary $\xi+\eta+\beta=0$, the dashed line is
the line $\xi\,(1-x)+b\,(-1-x)-\eta\,x=0$
for negative $x$, the arrows indicate the $\eta$
integration-paths for $(\xi+\eta+\beta)<0$
and for $(\xi+\eta+\beta)>0$
}\label{fig:figure_11}
\end{center}
\end{figure}

\begin{figure} [h!]
\begin{center}
\includegraphics[scale=0.5]{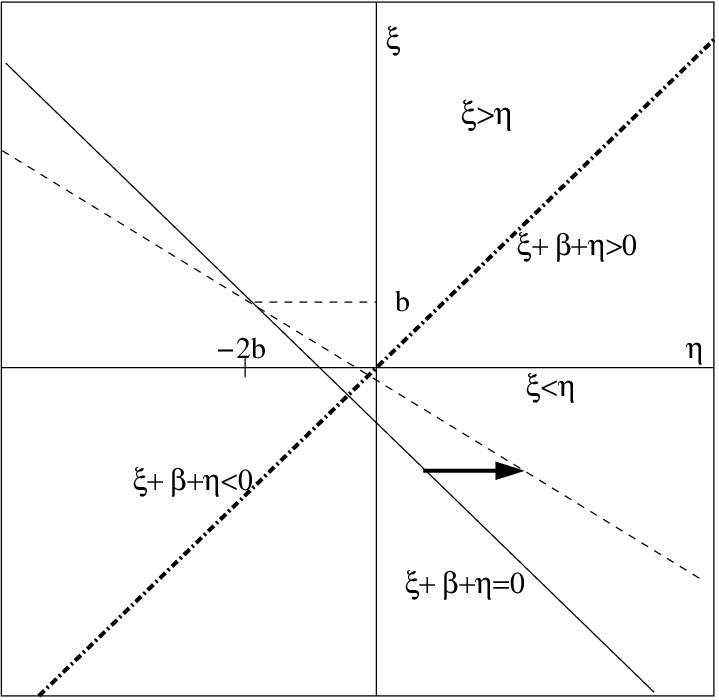}
\includegraphics[scale=0.5]{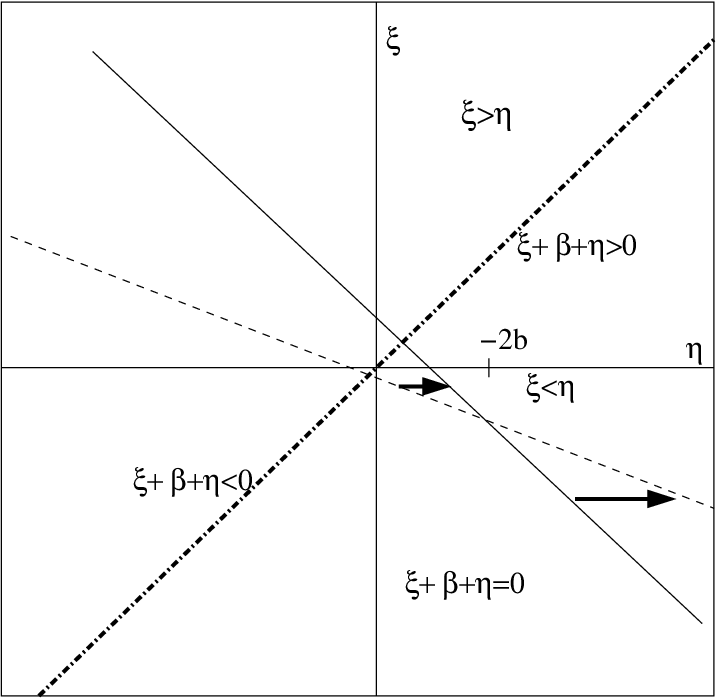}
\caption{\em Sector of the plane $(\eta,\xi)$ where $\xi+\eta+\beta>0$ or
$\xi+\eta+\beta<0$ and its boundary $\xi+\eta+\beta=0$, the dashed line is the line $\xi\,(1-x)+b\,(-1-x)-\eta\,x=0$
for negative $x$. The arrows indicate the $\eta$ integration-paths for $(\xi+\eta+\beta)<0$
and for $(\xi+\eta+\beta)>0\,$. One boundary of the integration regions is the line $\xi=\eta$
}\label{fig:figure_12}
\end{center}
\end{figure}

\begin{figure} [h!]
\begin{center}
\includegraphics[scale=0.5]{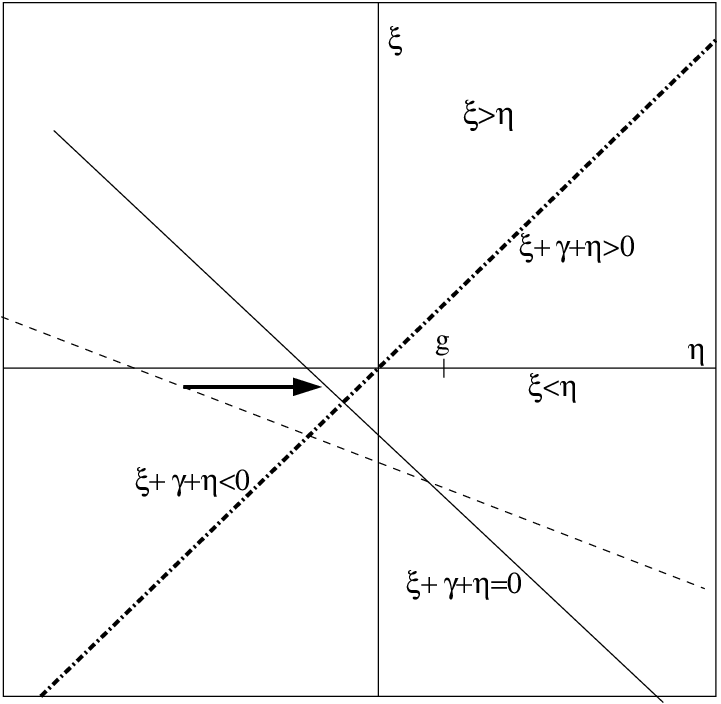}
\includegraphics[scale=0.5]{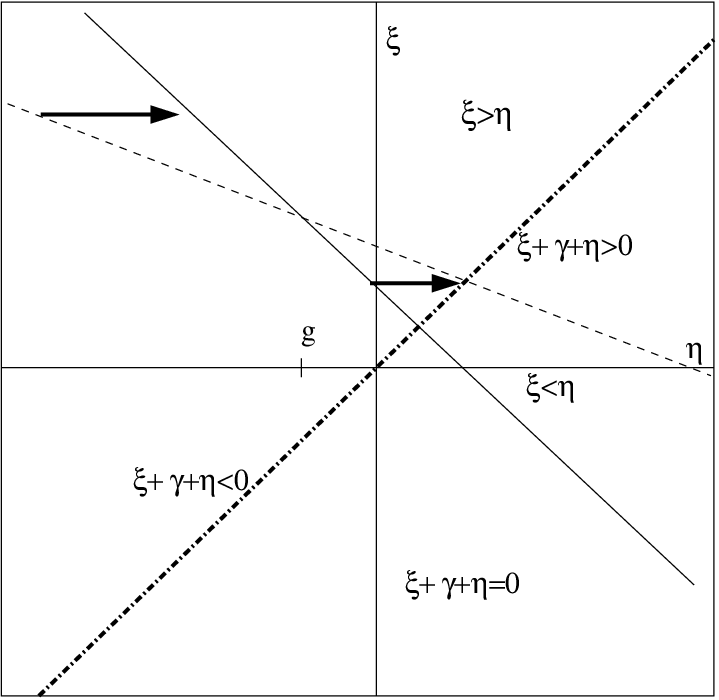}
\caption{\em Sector of the plane $(\eta,\xi)$ where
$\xi+\eta+\gamma>0$ or $\xi+\eta+\gamma<0$ and its boundary
$\xi+\eta+\gamma=0$, the dashed line is the line with
$\gamma=g$ and $\xi\,(1-x)+g\,(2-x)-\eta\,x=0$
for negative $x$. The arrows indicate the $\eta$
integration-paths for $(\xi+\eta+\gamma)<0$
and for $(\xi+\eta+\gamma)>0\,$, one boundary
of the integration regions is the dash-dotted line $\xi=\eta$
}\label{fig:figure_13}
\end{center}
\end{figure}

The PDF of $x_{g3}$ will be obtained differentiating the cumulative
probability to have the  $x_{g3}$ values less than a given value.
The following naming convention will be used in place
of the $\{x_i\}$:

\[x_1=\xi,\ \  x_2=\eta,\ \  x_3=\beta\ \   x_4=\gamma \ \ \]
giving
\[x_{g3}=\frac{\xi-\beta}{\xi+\eta+\beta} \ \ \ x_{g3}\leq 0.5\]
\[x_{g3}=\frac{2\gamma+\xi}{\gamma+\eta+\xi}\ \ \ x_{g3}>0.5\,. \]
In ref.~\cite{landi11}, we explored the conditions imposed by $x_{g3}$ to be lower than a given value $x$:
\begin{equation}
    \frac{\xi-\beta}{\xi+\eta+\beta}<x\,,
\end{equation}
with its two possibilities $\xi+\eta+\beta>0$ and $\xi+\eta+\beta<0$. In any case we have
two limiting planes:
\[ \xi(1-x)+\beta(-1-x)-\eta\, x=0\ \ \ \ \ \mathrm{and}\ \ \ \ \  \xi+\eta+\beta=0\,. \]
Let us see the case with $\beta>0$ and $\beta=b$. The traces of
the two planes are these of fig.~\ref{fig:figure_11}. The intersection point is $\{-2\,b,b\}$
and it changes with $\beta$.
Now we have to consider a change of algorithm when $\xi >\eta$ (with the
signals this implies $x_1>x_2$).
The plane $\xi,\eta$ must be separated in two parts along the
line $\xi=\eta$ below this line we have
the algorithm with the random variables $\beta,\xi,\eta$ with part
of the lines of fig.~\ref{fig:figure_11}. Above this line we have
the algorithm with the random variables $\gamma,\xi,\eta$
Let us explore the integration regions for $x<0$ and $\beta\neq 0$.
Figures~\ref{fig:figure_12} and~\ref{fig:figure_13} illustrate the conditions with $\beta\geq 0$ (left
side) and $\beta<0$ (right side) and similarly for the variable $\gamma$.

\subsection{A better reference system}

Observing the figures~\ref{fig:figure_11},~\ref{fig:figure_12}
and~\ref{fig:figure_13},  we see that the reference system based
on the two orthogonal lines $\xi+\eta+\beta=0$ and $\xi=\eta$ has
some special property. For any $x$, the integration regions have a
boundary on the line $\xi+\eta+\beta=0$, and the variation of $x$
implies a rotation of the dashed line around a fixed point (for fixed $\beta=b$).
For $x\geq 1/2$ a part of the plane $\xi,\eta$ is fully covered and the
integration regions change drastically. With this reference system
we have to distingue two range of $x$-values: $x\leq 1/2$ and $x>1/2$,
this is in fact the point in which we have a discontinuity.

\noindent
Let us write the form of the new variables (with $\beta$):
\begin{equation}
    \begin{aligned}
    &\eta'=\frac{\eta-\xi}{\sqrt{2}}  \ \ \ \ \ \ \ \
    &\eta=\frac{\xi'+\eta'}{\sqrt{2}}-\frac{\beta}{2}\\
    &\xi'=\frac{\eta+\xi+\beta}{\sqrt{2}} \ \ \ \ \ \
    &\xi=\frac{\xi'-\eta'}{\sqrt{2}}-\frac{\beta}{2}\\
    \end{aligned}
\end{equation}
and similarly for the side with $\gamma$:
\begin{equation}
    \begin{aligned}
    &\eta'=\frac{\eta-\xi}{\sqrt{2}}  \ \ \ \ \ \ \ \  &\eta=\frac{\xi'+\eta'}{\sqrt{2}}-\frac{\gamma}{2}\\
    &\xi'=\frac{\eta+\xi+\gamma}{\sqrt{2}} \ \ \ \ \ \  &\xi=\frac{\xi'-\eta'}{\sqrt{2}}-\frac{\gamma}{2}\\
    \end{aligned}
\end{equation}

\begin{figure} [h!]
\begin{center}
\includegraphics[scale=0.5]{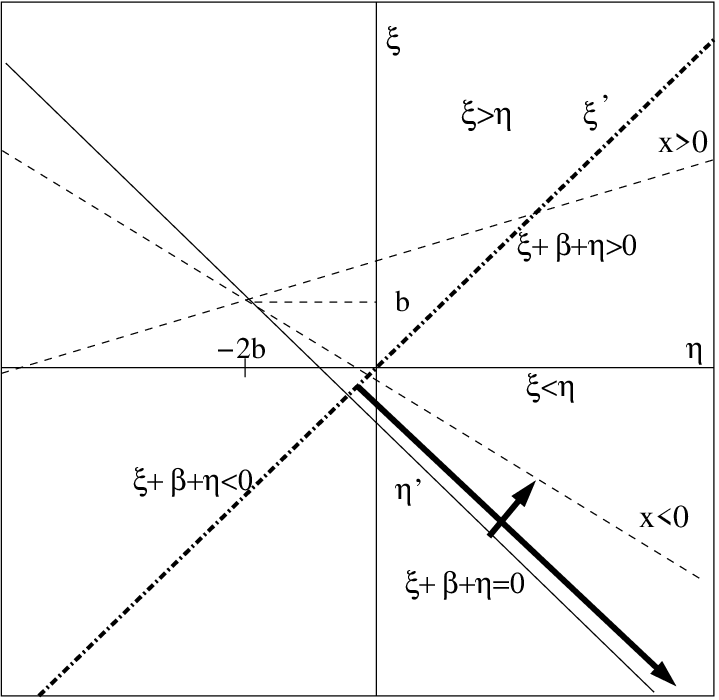}
\includegraphics[scale=0.5]{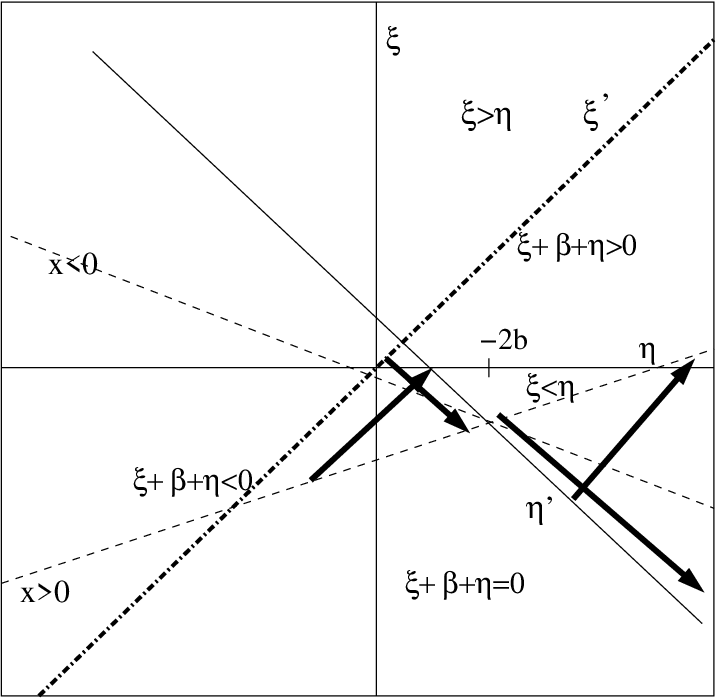}
\caption{\em The left figure is the case with $\beta>0$, the right side is for
$\beta<0$. The dashed lines are the line $\xi\,(1-x)+b\,(-1-x)-\eta\,x=0$ for
$x>0$ and $x<0$, the continuous line is the new axis $\eta'$ and the dash-dotted line
is the new axis $\xi'$.
For $x<1/2$ , the arrows indicate the $\eta'$ and $\xi'$ integration-paths for $(\xi+\eta+\beta)<0$
and for $(\xi+\eta+\beta)>0\,$
}\label{fig:figure_12a}
\end{center}
\end{figure}
The following integrals are those in the regions of figure~\ref{fig:figure_12a} for $\beta\geq 0$ and $\beta<0$.
\begin{equation}
    \begin{aligned}
    &F_1^{xg3}(x)=\\
    &\int_0^{+\infty}\mathrm{d}\beta P_3(\beta)\int_0^{+\infty}\mathrm{d}\eta'\int_0^{\frac{\sqrt{2}\eta'+3\beta}{\sqrt{2}(1-2x)}}
    \mathrm{d}\xi' P_2(\frac{\xi'+\eta'}{\sqrt{2}}-\frac{\beta}{2}) P_1(\frac{\xi'-\eta'}{\sqrt{2}}-\frac{\beta}{2})+\\
    &\int_{-\infty}^0\mathrm{d}\beta P_3(\beta)\int_0^{-\frac{3\beta}{\sqrt{2}}}\mathrm{d}\eta'\int_{\frac{\sqrt{2}\eta'+3\beta}{\sqrt{2}(1-2x)}}^0
    \mathrm{d}\xi'P_2(\frac{\xi'+\eta'}{\sqrt{2}}-\frac{\beta}{2}) P_1(\frac{\xi'-\eta'}{\sqrt{2}}-\frac{\beta}{2})+\\
    &\int_{-\infty}^0\mathrm{d}\beta P_3(\beta)\int_{-\frac{3\beta}{\sqrt{2}}}^{+\infty}\mathrm{d}\eta'\int_0^{\frac{\sqrt{2}\eta'+3\beta}{\sqrt{2}(1-2x)}}
    \mathrm{d}\xi'P_2(\frac{\xi'+\eta'}{\sqrt{2}}-\frac{\beta}{2}) P_1(\frac{\xi'-\eta'}{\sqrt{2}}-\frac{\beta}{2})\\
    \end{aligned}
\end{equation}
\begin{figure} [h!]
\begin{center}
\includegraphics[scale=0.5]{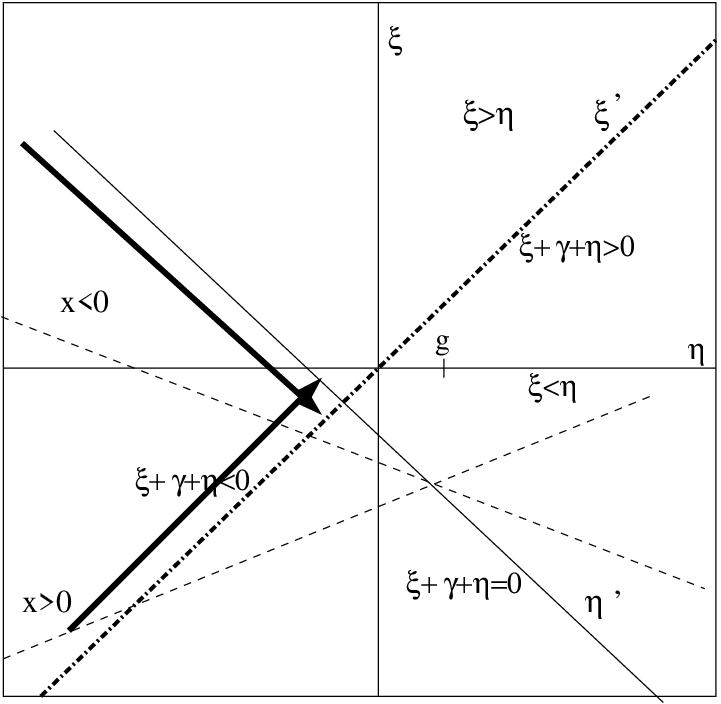}
\includegraphics[scale=0.5]{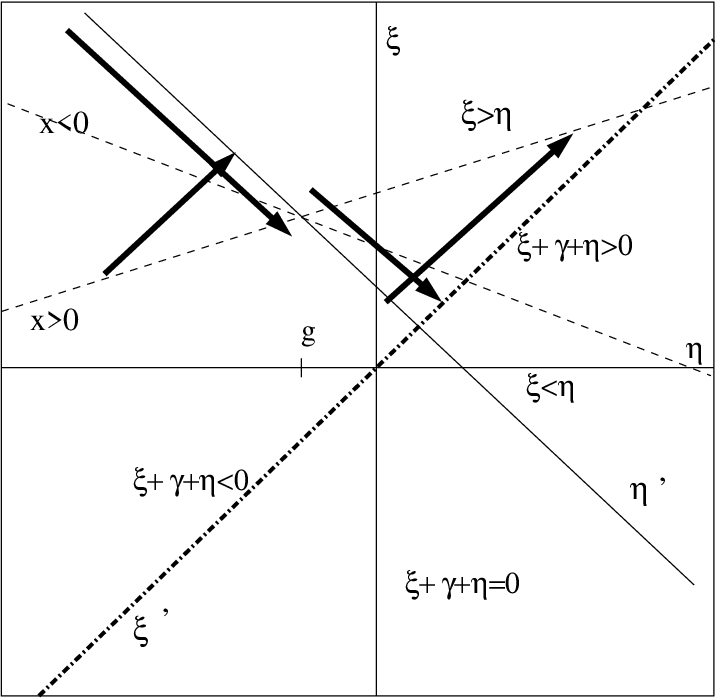}
\caption{\em Sector of the plane $(\eta,\xi)$ where $\xi+\eta+\gamma>0$ or
$\xi+\eta+\gamma<0$ and its boundary $\xi+\eta+\gamma=0$, the dashed line
is the line with $\gamma=g$ and $\xi\,(1-x)+g\,(2-x)-\eta\,x=0$.
For negative $x$, the arrows indicate the $\eta$ integration-paths for $(\xi+\eta+\gamma)<0$
and for $(\xi+\eta+\gamma)>0\,$ one of the boundary of the integration regions is the dash-dotted line $\xi=\eta$
}\label{fig:figure_13a}
\end{center}
\end{figure}
The integrals to the regions with $\eta'<0$ (figure~\ref{fig:figure_13a}) depend from the variable $\gamma$ and they are:
\begin{equation}
    \begin{aligned}
    &F_2^{xg3}(x)=\\
    &\int_0^{+\infty}\mathrm{d}\gamma P_4(\gamma)\int_{-\infty}^0\mathrm{d}\eta'\int_{\frac{\sqrt{2}\eta'-3\gamma}{\sqrt{2}(1-2x)}}^0
    \mathrm{d}\xi' P_2(\frac{\xi'+\eta'}{\sqrt{2}}-\frac{\gamma}{2}) P_1(\frac{\xi'-\eta'}{\sqrt{2}}-\frac{\gamma}{2})+\\
    &\int_{-\infty}^0\mathrm{d}\gamma P_4(\gamma)\int_{\frac{3\gamma}{\sqrt{2}}}^0\mathrm{d}\eta'\int_0^{\frac{\sqrt{2}\eta'-3\gamma}{\sqrt{2}(1-2x)}}
    \mathrm{d}\xi'P_2(\frac{\xi'+\eta'}{\sqrt{2}}-\frac{\gamma}{2}) P_1(\frac{\xi'-\eta'}{\sqrt{2}}-\frac{\gamma}{2})+\\
    &\int_{-\infty}^0\mathrm{d}\gamma P_4(\gamma)\int_{-\infty}^{\frac{3\gamma}{\sqrt{2}}}\mathrm{d}\eta'\int_0^{\frac{\sqrt{2}\eta'-3\gamma}{\sqrt{2}(1-2x)}}
    \mathrm{d}\xi'P_2(\frac{\xi'+\eta'}{\sqrt{2}}-\frac{\gamma}{2}) P_1(\frac{\xi'-\eta'}{\sqrt{2}}-\frac{\gamma}{2})\\
    \end{aligned}
\end{equation}
It is easy to verify that in the limit of $x\rightarrow$ $-\infty$ all the integrals in $\xi'$
converge to zero
and the cumulative function is zero, as it must be. Geometrically, this means that the integration
region below
the dashed lines of figures~\ref{fig:figure_12a} and~\ref{fig:figure_13a} disappear due to its
coincidence
with the $\xi'$ axis.

\subsection{Integrals for $x>1/2$}

Let us see the forms of the integrals for $x>1/2$, here some integrals are extended to fixed
regions of
the plane $\{\xi',\eta'\}$. The following figures~\ref{fig:figure_14a}
and~\ref{fig:figure_15a} illustrate
the integration regions for $x> 1/2$. The regions, that are unaffected by the $x$
variations, contribute to the cumulative distribution function, but are
irrelevant for the extraction
of the probability distribution. We will use them for a consistency check of the
cumulative distribution function with $x$ going to infinite.

\begin{figure} [h!]
\begin{center}
\includegraphics[scale=0.5]{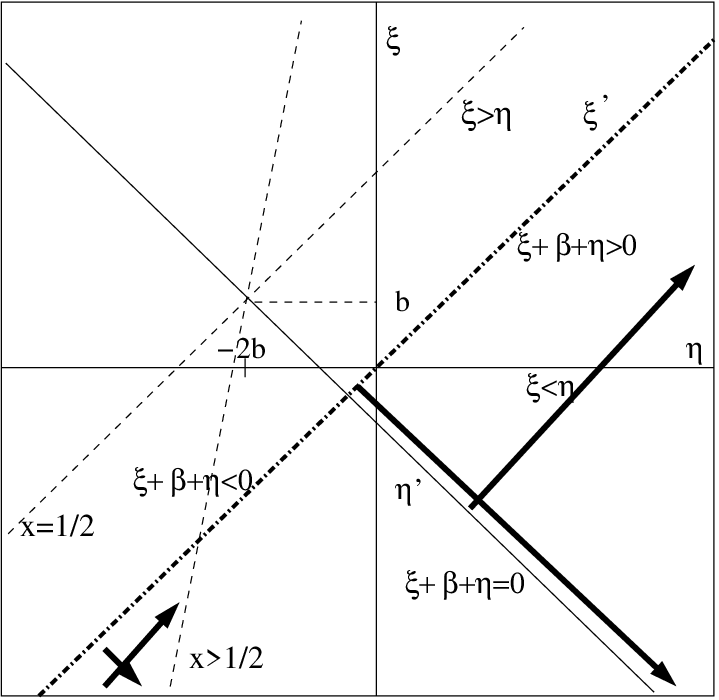}
\includegraphics[scale=0.5]{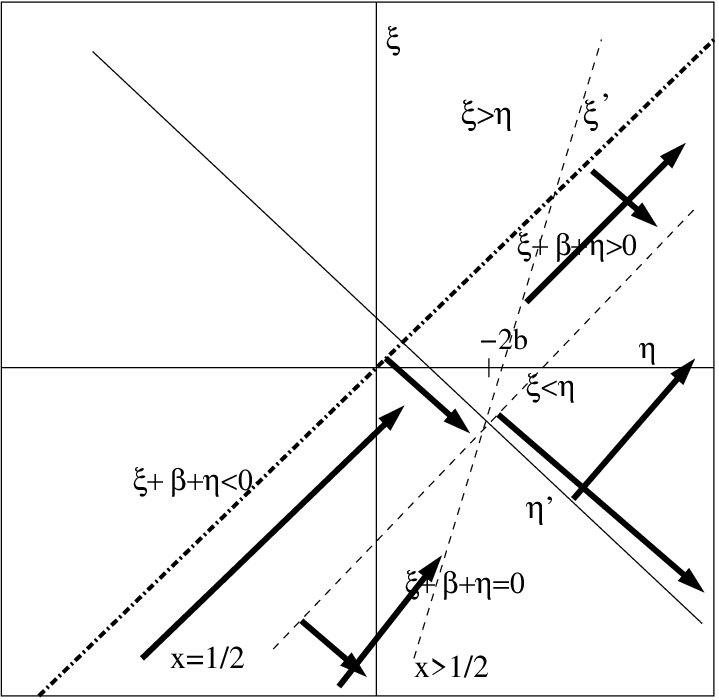}
\caption{\em The left figure is the case with $\beta>0$, the right side is for $\beta<0$. The dashed lines are the line $\xi\,(1-x)+b\,(-1-x)-\eta\,x=0$ for $x>0$ and $x<0$,the continuous line is the new axis $\eta'$ and the dash-dotted line
is the new axis $\xi'$.
For $x<1/2$ , the arrows indicate the $\eta'$ and $\xi'$ integration-paths for $(\xi+\eta+\beta)<0$
and for $(\xi+\eta+\beta)>0\,$
}\label{fig:figure_14a}
\end{center}
\end{figure}
Let us calculate the forms of the integrals for the random variable $\beta$. In this case $\eta>\xi$
and we are always below the line $\xi=\eta$ or for positive values of $\eta'$.
\begin{equation}
    \begin{aligned}
    &F_3^{xg3}(x)=\\
    &\int_0^{+\infty}\mathrm{d}\beta P_3(\beta)\int_0^{+\infty}\mathrm{d}\eta'\int_0^{+\infty}
    \mathrm{d}\xi' P_2(\frac{\xi'+\eta'}{\sqrt{2}}-\frac{\beta}{2}) P_1(\frac{\xi'-\eta'}{\sqrt{2}}-\frac{\beta}{2})+\\
    &\int_0^{+\infty}\mathrm{d}\beta P_3(\beta)\int_0^{+\infty}\mathrm{d}\eta'\int_{-\infty}^{\frac{\sqrt{2}\eta'+3\beta}{\sqrt{2}(1-2x)}}
    \mathrm{d}\xi' P_2(\frac{\xi'+\eta'}{\sqrt{2}}-\frac{\beta}{2}) P_1(\frac{\xi'-\eta'}{\sqrt{2}}-\frac{\beta}{2})+\\
    &\int_{-\infty}^0\mathrm{d}\beta P_3(\beta)\int_0^{-\frac{3\beta}{\sqrt{2}}}\mathrm{d}\eta'\int_{-\infty}^0
    \mathrm{d}\xi'P_2(\frac{\xi'+\eta'}{\sqrt{2}}-\frac{\beta}{2}) P_1(\frac{\xi'-\eta'}{\sqrt{2}}-\frac{\beta}{2})+\\
    &\int_{-\infty}^0\mathrm{d}\beta P_3(\beta)\int_{-\frac{3\beta}{\sqrt{2}}}^{+\infty}\mathrm{d}\eta'\int_{-\infty}^{\frac{\sqrt{2}\eta'+3\beta}{\sqrt{2}(1-2x)}}
    \mathrm{d}\xi'P_2(\frac{\xi'+\eta'}{\sqrt{2}}-\frac{\beta}{2}) P_1(\frac{\xi'-\eta'}{\sqrt{2}}-\frac{\beta}{2})+\\
    &\int_{-\infty}^0\mathrm{d}\beta P_3(\beta)\int_0^{-\frac{3\beta}{\sqrt{2}}}\mathrm{d}\eta'\int_{\frac{\sqrt{2}\eta'+3\beta}{\sqrt{2}(1-2x)}}^{+\infty}
    \mathrm{d}\xi'P_2(\frac{\xi'+\eta'}{\sqrt{2}}-\frac{\beta}{2}) P_1(\frac{\xi'-\eta'}{\sqrt{2}}-\frac{\beta}{2})+\\
    &\int_{-\infty}^0\mathrm{d}\beta P_3(\beta)\int_{-\frac{3\beta}{\sqrt{2}}}^{+\infty}\mathrm{d}\eta'\int_0^{+\infty}
    \mathrm{d}\xi'P_2(\frac{\xi'+\eta'}{\sqrt{2}}-\frac{\beta}{2}) P_1(\frac{\xi'-\eta'}{\sqrt{2}}-\frac{\beta}{2})\\
    \end{aligned}
\end{equation}
\begin{figure} [h!]
\begin{center}
\includegraphics[scale=0.5]{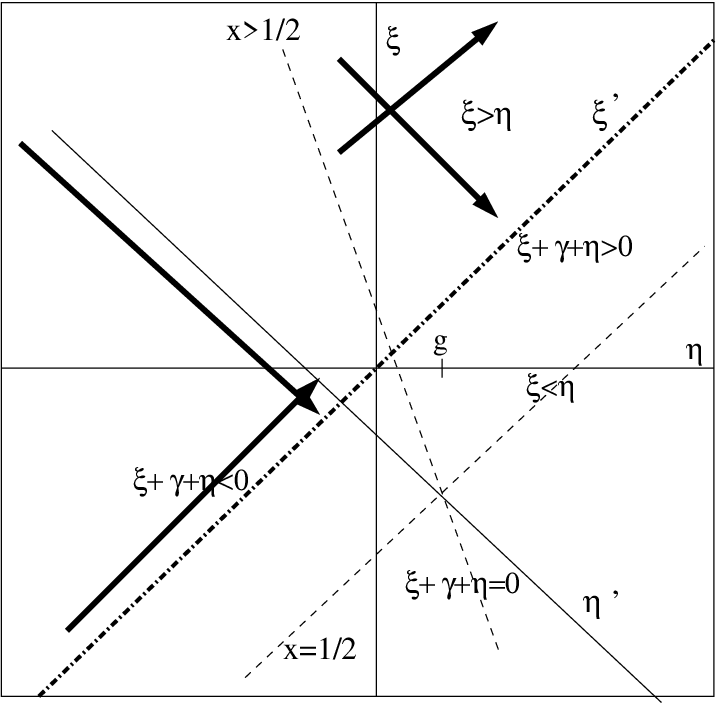}
\includegraphics[scale=0.5]{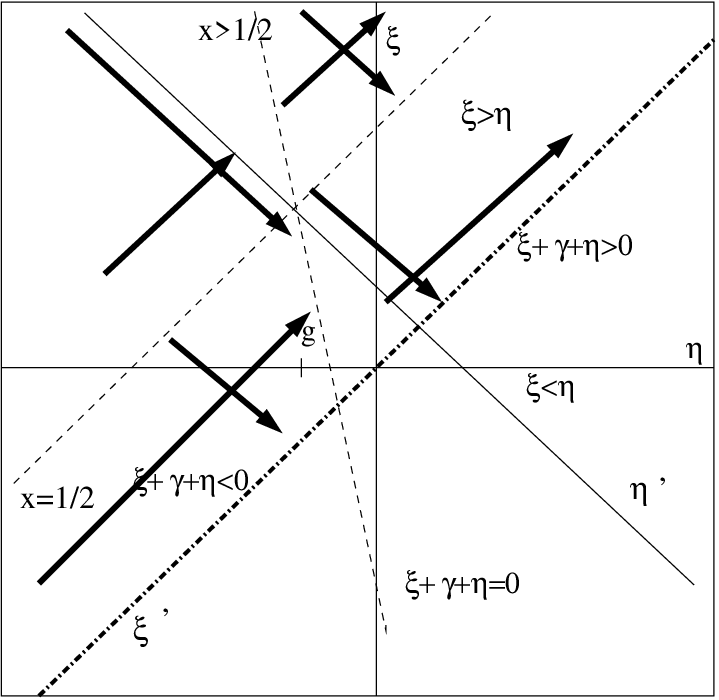}
\caption{\em Sector of the plane $(\eta,\xi)$ where $\xi+\eta+\gamma>0$ or
$\xi+\eta+\gamma<0$ and its boundary $\xi+\eta+\gamma=0$, the dashed line is the line with $\gamma=g$ $\xi\,(1-x)+g\,(2-x)-\eta\,x=0$
for negative $x$, the arrows indicate the $\eta$ integration-paths for $(\xi+\eta+\gamma)<0$
and for $(\xi+\eta+\gamma)>0\,$ one of the boundary of the integration regions is the dash-dotted line $\xi=\eta$
}\label{fig:figure_15a}
\end{center}
\end{figure}
The integrals with the random variable $\xi>\eta$ are the following:
\begin{equation}
    \begin{aligned}
    &F_4^{xg3}(x)=\\
    &\int_0^{+\infty}\mathrm{d}\gamma P_4(\gamma)\int_{-\infty}^0\mathrm{d}\eta'\int_{-\infty}^0
    \mathrm{d}\xi' P_2(\frac{\xi'+\eta'}{\sqrt{2}}-\frac{\gamma}{2}) P_1(\frac{\xi'-\eta'}{\sqrt{2}}-\frac{\gamma}{2})+\\
    &\int_0^{+\infty}\mathrm{d}\gamma P_4(\gamma)\int_{-\infty}^0\mathrm{d}\eta'\int_{\frac{\sqrt{2}\eta'-3\gamma}{\sqrt{2}(1-2x)}}^{+\infty}
    \mathrm{d}\xi' P_2(\frac{\xi'+\eta'}{\sqrt{2}}-\frac{\gamma}{2}) P_1(\frac{\xi'-\eta'}{\sqrt{2}}-\frac{\gamma}{2})+\\
    &\int_{-\infty}^0\mathrm{d}\gamma P_4(\gamma)\int_{-\infty}^{\frac{3\gamma}{\sqrt{2}}}\mathrm{d}\eta'\int_{-\infty}^0
    \mathrm{d}\xi'P_2(\frac{\xi'+\eta'}{\sqrt{2}}-\frac{\gamma}{2}) P_1(\frac{\xi'-\eta'}{\sqrt{2}}-\frac{\gamma}{2})+\\
    &\int_{-\infty}^0\mathrm{d}\gamma P_4(\gamma)\int_{\frac{3\gamma}{\sqrt{2}}}^0\mathrm{d}\eta'\int_{-\infty}^{\frac{\sqrt{2}\eta'-3\gamma}{\sqrt{2}(1-2x)}}
    \mathrm{d}\xi'P_2(\frac{\xi'+\eta'}{\sqrt{2}}-\frac{\gamma}{2}) P_1(\frac{\xi'-\eta'}{\sqrt{2}}-\frac{\gamma}{2})\\
    &\int_{-\infty}^0\mathrm{d}\gamma P_4(\gamma)\int_{-\infty}^{\frac{3\gamma}{\sqrt{2}}}\mathrm{d}\eta'\int_{\frac{\sqrt{2}\eta'-3\gamma}{\sqrt{2}(1-2x)}}^{+\infty}
    \mathrm{d}\xi'P_2(\frac{\xi'+\eta'}{\sqrt{2}}-\frac{\gamma}{2}) P_1(\frac{\xi'-\eta'}{\sqrt{2}}-\frac{\gamma}{2})+\\
    &\int_{-\infty}^0\mathrm{d}\gamma P_4(\gamma)\int_{\frac{3\gamma}{\sqrt{2}}}^0\mathrm{d}\eta'\int_0^{+\infty}
    \mathrm{d}\xi'P_2(\frac{\xi'+\eta'}{\sqrt{2}}-\frac{\gamma}{2}) P_1(\frac{\xi'-\eta'}{\sqrt{2}}-\frac{\gamma}{2})\\
    \end{aligned}
\end{equation}

To verify the consistency of the cumulative probability we must have:
\begin{equation}\label{eq:equation_14}
    \lim_{x\rightarrow+\infty}F^{xg3}(x)=\lim_{x\rightarrow+\infty}(F_3^{xg3}(x)+F_4^{xg3}(x))=1
\end{equation}
For $x\rightarrow +\infty$ the limit of the integrals that are expressed by $(\sqrt{2}\eta'-3\gamma)/(\sqrt{2}(1-2x))$ becomes zero and the first two of $F_4$ can be added giving
an integration on $\xi'$ from $-\infty$ to $+\infty$.The third and the forth produce an integral on
$\eta'$ from $-\infty$ to zero and an integral on $\xi'$ from $-\infty$ and 0. It is easy to see
the result (the integrations on the probability function of the random variable $\gamma$ and $\beta$, present respectively in the first and second integral, are omitted due to their trivial values equal to
one ):
\begin{equation}\label{eq:equation_16}%
\begin{aligned}
   &\lim_{x\rightarrow+\infty}F_3^{xg3}(x)=\int_0^{+\infty}\mathrm{d}\eta'
   \int_{-\infty}^{+\infty}\mathrm{d}\beta P_3(\beta)\int_{-\infty}^{+\infty}\mathrm{d}\xi'P_2(\frac{\xi'+\eta'}{\sqrt{2}}-\frac{\beta}{2}) P_1(\frac{\xi'-\eta'}{\sqrt{2}}-\frac{\beta}{2})\\
   &\lim_{x\rightarrow+\infty}F_4^{xg3}(x)=\int_{-\infty}^0\mathrm{d}\eta'
   \int_{-\infty}^{+\infty}\mathrm{d}\gamma P_4(\gamma)\int_{-\infty}^{+\infty}\mathrm{d}\xi'P_2(\frac{\xi'+\eta'}{\sqrt{2}}-\frac{\gamma}{2}) P_1(\frac{\xi'-\eta'}{\sqrt{2}}-\frac{\gamma}{2})\\
   \end{aligned}
\end{equation}

The observation of fig.~\ref{fig:figure_14a} and fig.~\ref{fig:figure_15a} for $x\rightarrow+\infty$
shows that the integration regions are independent from the parameters $\beta$ and $\gamma$ and the
variable $\xi'$ can be defined for $\beta=b=0$ and $\gamma=g=0$. Equation~\ref{eq:equation_16} becomes:

\begin{equation}
\begin{aligned}
   &\lim_{x\rightarrow+\infty}F_3^{xg3}(x)=\int_0^{+\infty}\mathrm{d}\eta'
   \int_{-\infty}^{+\infty}\mathrm{d}\xi'P_2(\frac{\xi'+\eta'}{\sqrt{2}}) P_1(\frac{\xi'-\eta'}{\sqrt{2}})\\
   &\lim_{x\rightarrow+\infty}F_4^{xg3}(x)=\int_{-\infty}^0\mathrm{d}\eta'
   \int_{-\infty}^{+\infty}\mathrm{d}\xi'P_2(\frac{\xi'+\eta'}{\sqrt{2}}) P_1(\frac{\xi'-\eta'}{\sqrt{2}})\\
\end{aligned}
\end{equation}
The integrals on $\beta$ and $\gamma$ are one for their normalization. The two remaining integrals sum
together and all the space of the variables $\xi$ and $\eta$ is covered and their normalization gives eq.~\ref{eq:equation_14}.

\subsection{COG$_3$ probability density function}

The differentiation of $F^{xg3}(x)$ in $x$ gives the probability density function $P_d^{xg3}(x)$. The
forms of $F^{xg3}(x)$ for $x>1/2$ and $x\leq 1/2$ give an identical expression upon differentiation.
\begin{equation}\label{eq:equation_17}
    \begin{aligned}
    &P_d^{xg3}(x)=\frac{\mathrm{d}F^{xg3}(x)}{\mathrm{d}x}=\\
    &\int_0^{+\infty}\mathrm{d}\beta P_3(\beta)\int_0^{+\infty}\mathrm{d}\eta' P_2(\frac{\sqrt{2}(1-x)\eta'+\beta(1+x)}{1-2x})P_1(\frac{\sqrt{2}x\eta'+\beta(1+x)}{1-2x})
    \frac{2\eta'+3\sqrt{2}\beta}{(1-2x)^2}\\
    &+\int_{-\infty}^0\mathrm{d}\beta P_3(\beta)\int_{-\frac{3\beta}{\sqrt{2}}}^{+\infty}\mathrm{d}\eta' P_2(\frac{\sqrt{2}(1-x)\eta'+\beta(1+x)}{1-2x})P_1(\frac{\sqrt{2}x\eta'+\beta(1+x)}{1-2x})
    \frac{2\eta'+3\sqrt{2}\beta}{(1-2x)^2}\\
    &-\int_{-\infty}^0\mathrm{d}\beta P_3(\beta)\int_0^{-\frac{3\beta}{\sqrt{2}}}\mathrm{d}\eta' P_2(\frac{\sqrt{2}(1-x)\eta'+\beta(1+x)}{1-2x})P_1(\frac{\sqrt{2}x\eta'+\beta(1+x)}{1-2x})
    \frac{2\eta'+3\sqrt{2}\beta}{(1-2x)^2}\\
    &-\int_0^{+\infty}\mathrm{d}\gamma P_4(\gamma)\int_{-\infty}^0\mathrm{d}\eta'P_2(\frac{\sqrt{2}(1-x)\eta'-\gamma(2-x)}{1-2x})
    P_1(\frac{\sqrt{2}x\eta'-\gamma(2-x)}{1-2x})
    \frac{2\eta'-3\sqrt{2}\gamma}{(1-2x)^2}\\
    &+\int_{-\infty}^0\mathrm{d}\gamma P_4(\gamma)\int_{\frac{3\gamma}{\sqrt{2}}}^0\mathrm{d}\eta'P_2(\frac{\sqrt{2}(1-x)\eta'-\gamma(2-x)}{1-2x})
    P_1(\frac{\sqrt{2}x\eta'-\gamma(2-x)}{1-2x})
    \frac{2\eta'-3\sqrt{2}\gamma}{(1-2x)^2}\\
    &+\int_{-\infty}^0\mathrm{d}\gamma P_4(\gamma)\int_{-\infty}^{\frac{3\gamma}{\sqrt{2}}}\mathrm{d}\eta'P_2(\frac{\sqrt{2}(1-x)\eta'-\gamma(2-x)}{1-2x})
    P_1(\frac{\sqrt{2}x\eta'-\gamma(2-x)}{1-2x})
    \frac{2\eta'-3\sqrt{2}\gamma}{(1-2x)^2}\\
    \end{aligned}
\end{equation}
Equation~\ref{eq:equation_17} can be recast in the form (explicitly positive):
\begin{equation}\label{eq:equation_18}
    \begin{aligned}
    &P_d^{xg3}(x)=\frac{\mathrm{d}F^{xg3}(x)}{\mathrm{d}x}=\\
    &\int_{-\infty}^{+\infty}\mathrm{d}\beta P_3(\beta)\int_0^{+\infty}\mathrm{d}\eta' P_2(\frac{\sqrt{2}(1-x)\eta'+\beta(1+x)}{1-2x})P_1(\frac{\sqrt{2}x\eta'+\beta(1+x)}{1-2x})
    \frac{\mathrm{abs}(2\eta'+3\sqrt{2}\beta)}{(1-2x)^2}+\\
    &\int_{-\infty}^{+\infty}\mathrm{d}\gamma P_4(\gamma)\int_{-\infty}^0\mathrm{d}\eta'P_2(\frac{\sqrt{2}(1-x)\eta'-\gamma(2-x)}{1-2x})
    P_1(\frac{\sqrt{2}x\eta'-\gamma(2-x)}{1-2x})
    \frac{\mathrm{abs}(2\eta'-3\sqrt{2}\gamma)}{(1-2x)^2}\\
    \end{aligned}
\end{equation}
Equation~\ref{eq:equation_18} is general for any $x$ and any distribution of the four
random variables. With a set of appropriate coordinate transformations the
integrals of equation 2 of ref.~\cite{landi12} can be recast in this form.
The denominators $(1-2x)^2$ allow to introduce a Dirac $\delta$ approximation
around $x\approx 1/2$ as done in ref.~\cite{landi06} for the two strip COG
around $x\approx 0$. The Cauchy-Agnesi tails are evident in~\ref{eq:equation_18}.
Renaming the random variables and the due change of $x$, eq.~\ref{eq:equation_18}
can be used around the gap/discontinuity $x\approx -1/2$.

\section{Probability distributions for the four strip COG (COG$_4$) }

As discussed in ref.~\cite{landi01}, the histograms of the COG algorithms
with an even number of strips (and large signal distributions) have
gaps/discontinuities around $x\approx 0$. Thus, as for the two strip COG,
we have to separately study the addition of the right strip and the addition
of the left strip.

\subsection{COG$_4$ probability distributions with only right strip or left strip}

Now we have to work in a four dimensional space. We will consider the two strips around
the strip with the maximum signal and the forth strip to the right of these three. Let us
recall our unusual strip ordering for the five strips required here (from left to right): $<5>$, $<3>$,
$<2>$, $<1>$, $<4>$. With the convention on
the names of the strip signals:

\[x_5=\psi,\ \ \ x_3=\beta, \ \ \  x_2=\eta,\ \ \  x_1=\xi,\ \ \  x_4=\gamma\,.\]

\begin{figure} [h!]
\begin{center}
\includegraphics[scale=0.5]{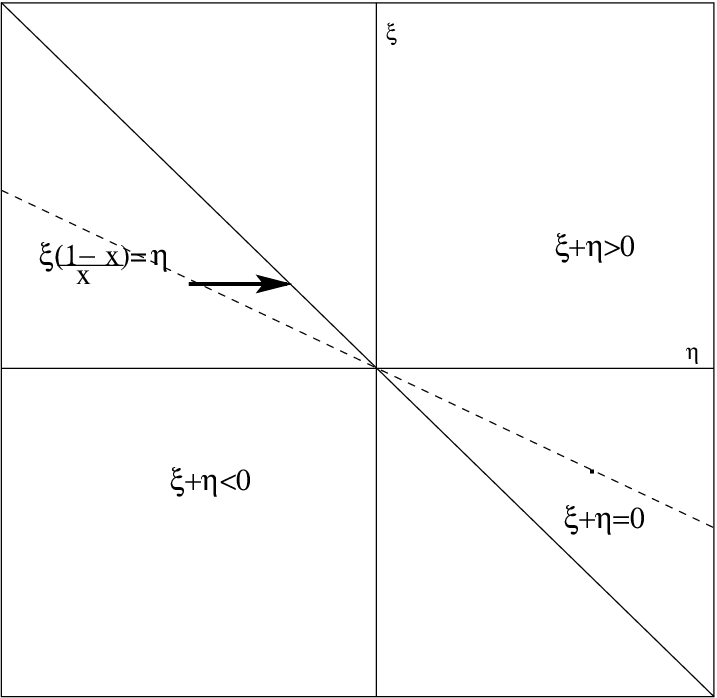}
\caption{\em Sector of the plane $(\eta,\xi)$ with $\beta=\gamma=0$ where $\xi+\eta>0$ or
$\xi+\eta<0$ and its boundary $\xi+\eta=0$, the dashed line is the line $\xi(1-x)/x=\eta$
for negative $x$, the arrow indicates the $\eta$ integration-path for $(\xi+\eta)<0$
}\label{fig:figure_1}
\end{center}
\end{figure}

As always the maximum random variable is $x_2=\eta$ and the center of the strip $<2>$ is the origin of the axis
for the COG$_4$ algorithm. In this reference system $x_{g4}$ is:

\[\ \ x_{g4}=\frac{\xi-\beta+2\gamma}{\xi+\eta+\beta+\gamma} \]
For $\beta=0$, $\gamma=0$ and $\xi+\eta>0$ we have the plot of fig.~\ref{fig:figure_1}
in general we have to explore the condition:
\begin{equation}
    \frac{\xi-\beta+2\gamma}{\xi+\eta+\beta+\gamma}<x.
\end{equation}
We have two possibilities $\xi+\eta+\beta+\gamma>0$ and $\xi+\eta+\beta+\gamma<0$, in any case we have
two limiting planes:
\[ \xi(1-x)+\beta(-1-x)-\eta\, x+\gamma(2-x)=0\ \ \ \ \mathrm{and}\ \ \ \  \xi+\eta+\beta+\gamma=0 \]
for $x<0$ the traces of these two planes in the plane $\eta,\xi$
are these of figure.~\ref{fig:figure_1}.

\begin{figure} [h!]
\begin{center}
\includegraphics[scale=0.5]{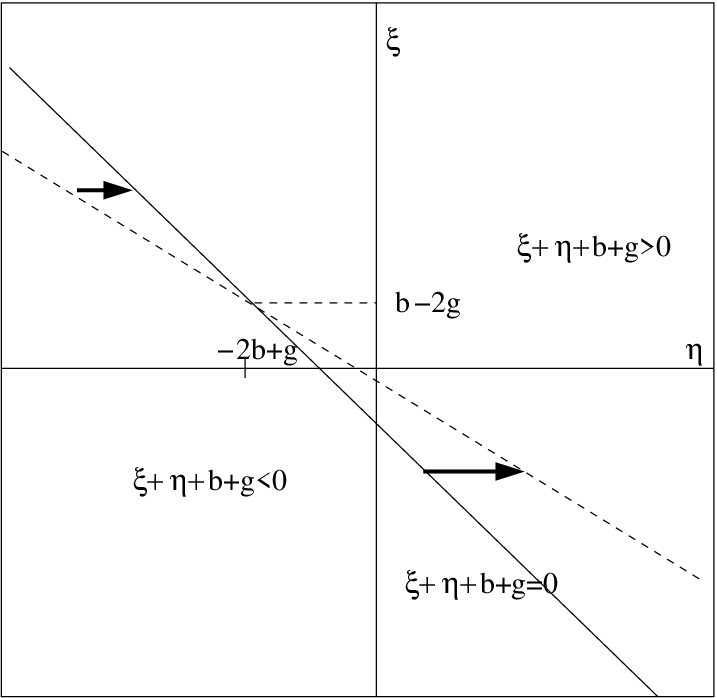}
\caption{\em Sector of the plane $(\eta,\xi)$ for a fixed value of $\beta=b$ and $\gamma=g$ where $\xi+\eta+b+g>0$ or
$\xi+\eta+b+g<0$ and its boundary $\xi+\eta+b+g=0$, the dashed line has the equation
 $\xi\,(1-x)+b\,(-1-x)-\eta\,x+g\,(2-x)=0$
for negative $x$, the arrows indicate the $\eta$ integration-paths for $(\xi+\eta+b+g)<0$
and for $(\xi+\eta+b+g)>0$. The intersection point of the two boundary lines is $\{-2b+g,b-2g\}$
}\label{fig:figure_11b}
\end{center}
\end{figure}

Let us see the case
with $\beta>0$, $\gamma>0$ and $\beta=b$ and $\gamma=g$.
We have two conditions that define the allowed regions of the
random variables, for this it will be possible to consider two
of the four variables completely unconstrained. The constraint will
be applied to the variables $\xi$ and $\eta$ as done always.
The variables $\xi$ and $\eta$ must be contained within the
two planes of fig.~\ref{fig:figure_11b}. The intersection point is $\{-2\,b+g,b-2g\}$
and now it follows the values of $\beta$ and $\gamma$.
Evidently, the only change in moving from $\beta>0$ and $\gamma>0$ to
$\beta<0$ and $\gamma<0$ or any other combination of signs is a
shift of the intersection point, and each line of the plot moves parallel to itself.
As for the three strip case $F_1^{xg4}(x)$ becomes ($x<0$):
\begin{equation}\label{eq:equation_100}
\begin{aligned}
    F_1^{xg4}(x)=&\int_{-\infty}^{\infty}\mathrm{d}\gamma\,P_4(\gamma)\int_{-\infty}^\infty\,\mathrm{d}\beta\,P_3(\beta)
    \int_{\beta-2\gamma}^\infty\,\mathrm{d}\xi\,P_1(\xi)
    \int_{\xi\frac{(1-x)}{x}+\beta\frac{(-1-x)}{x}+\gamma\frac{2-x}{x}}^{-\xi-\beta-\gamma}P_2(\eta)\mathrm{d}\eta+\\
    &\int_{-\infty}^{\infty}\,\mathrm{d}\gamma\,P_4(\gamma)\int_{-\infty}^\infty\,\mathrm{d}\beta\,P_3(\beta)\int_{-\infty}^{\beta-2\gamma},\mathrm{d}\xi\,P_1(\xi)
    \int_{-\xi-\beta-\gamma}^{\xi\frac{(1-x)}{x}+\beta\frac{(-1-x)}{x}+\gamma\frac{2-x}{x}}P_2(\eta)\mathrm{d}\eta\\
\end{aligned}
\end{equation}

\begin{figure} [h!]
\begin{center}
\includegraphics[scale=0.5]{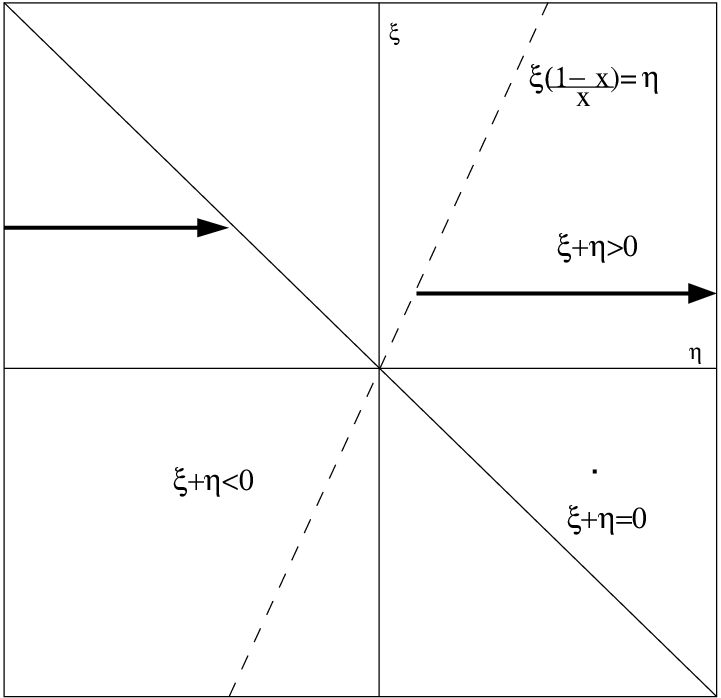}
\caption{\em Sector of the plane $(\eta,\xi)$ with $\beta=0$ and $\gamma=0$ where $\xi+\eta>0$ or
$\xi+\eta<0$ and its boundary $\xi+\eta=0$, the dashed line is the line $\xi(1-x)/x=\eta$ for
positive $x$. The arrows indicate the integration regions and these must cover positive and negative
values of $\xi$  }\label{fig:figure_12b}
\end{center}
\end{figure}

For $x\geq 0$ the traces of the two planes for $\beta=0$ and $\gamma=0$ are illustrated in
fig.~\ref{fig:figure_12b}.
With $\beta\neq 0$ and $\gamma\neq 0$, the traces of the two planes
become these of fig.~\ref{fig:figure_13b}. The arrows indicate the integration paths.
The intersection of the two lines are always in the point $\{-2b+g,b-2g\}$ and the $F_2^{xg4}(x)$ becomes:
\begin{figure} [h!]
\begin{center}
\includegraphics[scale=0.5]{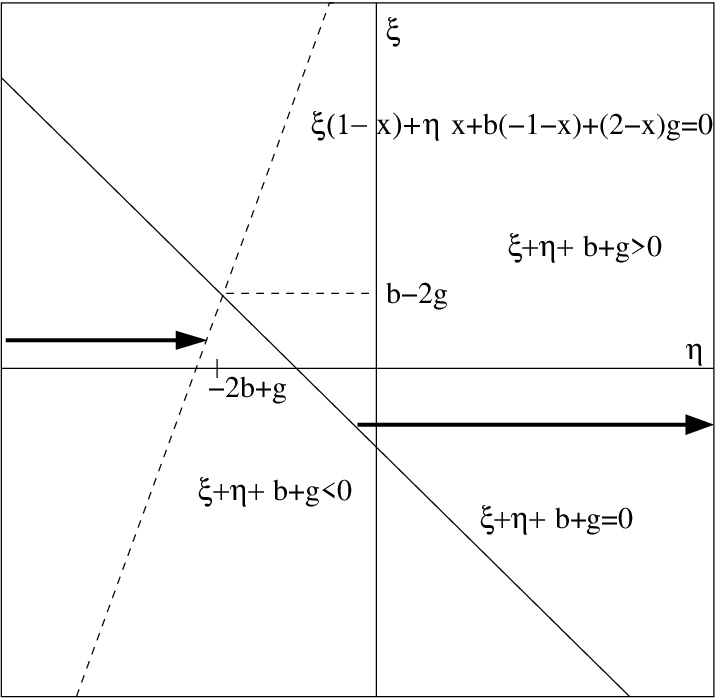}
\caption{\em Sector of the plane $(\eta,\xi)$ where $\xi+\eta+b+g>0$ or
$\xi+\eta+b+g<0$ and its boundary $\xi+\eta+b+g=0$, the dashed line is the line $\xi(1-x)-x\eta+b(-1-x)+g(2-x)=0$ for
$x>0$. The arrows indicate the integration regions and these must cover positive and negative
values of $\xi$  }\label{fig:figure_13b}
\end{center}
\end{figure}
\begin{equation}\label{eq:equation_110}
\begin{aligned}
    F_2^{xg4}(x)=&\int_{-\infty}^{\infty}\,\mathrm{d}\gamma\,P_4(\gamma)\int_{-\infty}^\infty\,\mathrm{d}\beta\,P_3(\beta)
    \int_{-\infty}^{\beta-2\gamma}\,\mathrm{d}\xi\,P_1(\xi)
    \int_{-\xi-\beta-\gamma}^{\infty}P_2(\eta)\mathrm{d}\eta+\\
    &\int_{-\infty}^{\infty}\,\mathrm{d}\gamma\,P_4(\gamma)\int_{-\infty}^\infty\,\mathrm{d}\beta\,P_3(\beta)
    \int_{\beta-2\gamma}^{\infty}\,\mathrm{d}\xi\,P_1(\xi)
    \int_{\xi\frac{(1-x)}{x}+\beta\frac{(-1-x)}{x}+\gamma\frac{2-x}{x}}^{+\infty}P_2(\eta)\mathrm{d}\eta+\\
    &\int_{-\infty}^{\infty}\,\mathrm{d}\gamma\,P_4(\gamma)\int_{-\infty}^\infty\,\mathrm{d}\beta\,P_3(\beta)
    \int_{-\infty}^{\beta-2\gamma},\mathrm{d}\xi\,P_1(\xi)
    \int_{-\infty}^{\xi\frac{(1-x)}{x}+\beta\frac{(-1-x)}{x}+\gamma\frac{2-x}{x}}P_2(\eta)\mathrm{d}\eta\\
    &\int_{-\infty}^{\infty}\,\mathrm{d}\gamma\,P_4(\gamma)\int_{-\infty}^\infty\,\mathrm{d}\beta\,P_3(\beta)
    \int_{\beta-2\gamma}^{+\infty},\mathrm{d}\xi\,P_1(\xi)
    \int_{-\infty}^{-\xi-\beta-\gamma}P_2(\eta)\mathrm{d}\eta
\end{aligned}
\end{equation}
As for $x_{g3}$, it is easy to prove that $\lim_{x\rightarrow -\infty}F_1^{xg4}(x)=0$ and
$\lim_{x\rightarrow +\infty}F_2^{xg4}(x)=1$. In fact the first limit is easy given that
$\lim_{x\rightarrow \pm\infty}(\pm 1-x)/x=-1$ and $\lim_{x\rightarrow \pm\infty}(2-x)/x=-1$ . With this position eq.~\ref{eq:equation_100} has
the limits of the last integrals identical, and the integrals are zero.

For $x\rightarrow +\infty$ the integrals of eq.~\ref{eq:equation_110} become:
\begin{equation}\label{eq:equation_120z}
\begin{aligned}
    F_2^{xg4}(+\infty)=&\int_{-\infty}^{\infty}\,\mathrm{d}\gamma\,P_4(\gamma)
    \int_{-\infty}^\infty\,\mathrm{d}\beta\,P_3(\beta)\int_{-\infty}^{\beta-2\gamma}\,\mathrm{d}\xi\,P_1(\xi)
    \int_{-\xi-\beta-\gamma}^{\infty}P_2(\eta)\mathrm{d}\eta+\\
    &\int_{-\infty}^{\infty}\,\mathrm{d}\gamma\,P_4(\gamma)
    \int_{-\infty}^\infty\,\mathrm{d}\beta\,P_3(\beta)\int_{\beta-2\gamma}^{\infty}\,\mathrm{d}\xi\,P_1(\xi)
    \int_{-\xi-\beta-\gamma}^{+\infty}P_2(\eta)\mathrm{d}\eta+\\
    &\int_{-\infty}^{\infty}\,\mathrm{d}\gamma\,P_4(\gamma)
    \int_{-\infty}^\infty\,\mathrm{d}\beta\,P_3(\beta)\int_{-\infty}^{\beta-2\gamma},\mathrm{d}\xi\,P_1(\xi)
    \int_{-\infty}^{-\xi-\beta-\gamma}P_2(\eta)\mathrm{d}\eta\\
    &\int_{-\infty}^{\infty}\,\mathrm{d}\gamma\,P_4(\gamma)
    \int_{-\infty}^\infty\,\mathrm{d}\beta\,P_3(\beta)\int_{\beta-2\gamma}^{+\infty},\mathrm{d}\xi\,P_1(\xi)
    \int_{-\infty}^{-\xi-\beta-\gamma}P_2(\eta)\mathrm{d}\eta
\end{aligned}
\end{equation}
The first and third integral have identical integration limits in the variables $\gamma$, $\beta$ and
$\xi$,  and the sum of the last integrals produces the normalization of the
probability distribution $P_2(\eta)$. Identically for the
second and forth integrals. The two remaining integrals add completing the normalization of the
probability $P_1(\xi)$ that is multiplied by the normalization of $P_3(\beta)$ and $P_4(\gamma)$
giving 1.

Now, after this consistency check, we can extract the probability $P_{xg3}(x)$
differentiating $F_1^{xg4}(x)$ and $F_2^{xg4}(x)$ respect to $x$. The result is:
\begin{equation}\label{eq:equation_120}
\begin{aligned}
    &P_{xg4}(x)=\\
    &\frac{1}{x^2}\Big[\int_{-\infty}^{\infty}\mathrm{d}\gamma P_4(\gamma)
    \int_{-\infty}^{+\infty}\mathrm{d}\beta P_3(\beta)\int_{\beta-2\gamma}^{+\infty}\mathrm{d}\xi
    P_1(\xi)P_2(\xi\frac{1-x}{x}+\beta\frac{-1-x}{x}+\gamma\frac{2-x}{x})(-\beta+\xi+2\gamma)+\\
    &\int_{-\infty}^{\infty}\,\mathrm{d}\gamma\,P_4(\gamma)
    \int_{-\infty}^{+\infty}\mathrm{d}\beta\,P_3(\beta)\int_{-\infty}^{\beta-2\gamma}\mathrm{d}\xi\,
    P_1(\xi)P_2(\xi\frac{1-x}{x}+
    \beta\frac{-1-x}{x}+\gamma\frac{2-x}{x})(\beta-\xi-2\gamma)\Big]
\end{aligned}
\end{equation}
With the transformation $\delta=\xi-\beta+2\gamma$, eq.~\ref{eq:equation_120} becomes:
\begin{equation}\label{eq:equation_130}
\begin{aligned}
    P_{xg4}(x)=&\frac{1}{x^2}\Big[\int_{-\infty}^{\infty}\,\mathrm{d}\gamma\,P_4(\gamma)
    \int_{-\infty}^{+\infty}\mathrm{d}\beta\,P_3(\beta)\int_0^{+\infty}\mathrm{d}\delta\,
    P_1(\delta+\beta-2\gamma)P_2(\delta\frac{1-x}{x}-2\beta+\gamma)\delta-\\
    &\int_{-\infty}^{\infty}\,\mathrm{d}\gamma\,P_4(\gamma)
    \int_{-\infty}^{+\infty}\mathrm{d}\beta\,P_3(\beta)\int_{-\infty}^{0}\mathrm{d}\delta\,P_1(\delta+\beta-2\gamma)P_2(\delta\frac{1-x}{x}
    -2\beta+\gamma)\delta\Big]
\end{aligned}
\end{equation}
or better:
\begin{equation}\label{eq:equation_130b}
    P_{xg4}(x)=\frac{1}{x^2}\Big[\int_{-\infty}^{\infty}\,\mathrm{d}\gamma\,P_4(\gamma)
    \int_{-\infty}^{+\infty}\mathrm{d}\beta\,P_3(\beta)\int_{-\infty}^{+\infty}\mathrm{d}\delta\,
    P_1(\delta+\beta-2\gamma)P_2(\delta\frac{1-x}{x}-2\beta+\gamma)\mathrm{abs}(\delta)\Big]
\end{equation}
This can be recast in a form more appropriate for its use with gaussian probability
distributions:
\begin{equation}\label{eq:equation_130a}
    P_{xg4}(x)=\frac{1}{x^2}\Big[
    \int_{-\infty}^{+\infty}\mathrm{d}\delta\,\mathrm{abs}(\delta)\,
    \int_{-\infty}^{\infty}\,\mathrm{d}\gamma\,P_4(\gamma)\int_{-\infty}^{+\infty}\mathrm{d}\beta\,P_3(\beta)
    P_1(\delta+\beta-2\gamma)P_2(\delta\frac{1-x}{x}-2\beta+\gamma)\Big]
\end{equation}


Most of the approach for the right strip can be reused for the left strip, in this
case the COG$_4$ algorithm is given:

\[x_5=\psi,\ \ x_3=\beta, \ \ x_2=\eta,\ \ x_1=\xi,\ \   x_4=\gamma\ \ \ \ \
x_{g4l}=\frac{\xi-\beta-2\psi}{\xi+\eta+\beta+\psi}1,. \]

\noindent
Our condition is:

\begin{equation}
    \frac{\xi-\beta-2\psi}{\xi+\eta+\beta+\psi}<x\,.
\end{equation}
This condition limits the random variables to be
contained in regions bounded by the two planes:
\[ \xi(1-x)+\beta(-1-x)-\eta\, x+\psi(-2-x)=0\ \ \ \ \mathrm{and}\ \ \ \  \xi+\eta+\beta+\psi=0\,. \]
The two planes intersect in the $\eta,\xi$-plane in the point $-2b-3p,b+2p$
if $\beta=b$ and $\psi=p$. In any case the developments are very
similar to these performed for the right-strip case. The
random variable $\psi$ has probability distribution $P_5(\psi)$. Hence,
$F_2^{xg4}(x)$ becomes ($x<0$):
\begin{equation}\label{eq:equation_150}
\begin{aligned}
    F_3^{xg4}(x)=&\int_{-\infty}^{\infty}\mathrm{d}\psi\,P_5(\psi)\int_{-\infty}^\infty\,\mathrm{d}\beta\,P_3(\beta)
    \int_{\beta+2\psi}^\infty\,\mathrm{d}\xi\,P_1(\xi)
    \int_{\xi\frac{(1-x)}{x}+\beta\frac{(-1-x)}{x}+\psi\frac{-2-x}{x}}^{-\xi-\beta-\psi}P_2(\eta)\mathrm{d}\eta+\\
    &\int_{-\infty}^{\infty}\,\mathrm{d}\psi\,P_5(\psi)\int_{-\infty}^\infty\,\mathrm{d}\beta\,P_3(\beta)\int_{-\infty}^{\beta+2\psi},\mathrm{d}\xi\,P_1(\xi)
    \int_{-\xi-\beta-\psi}^{\xi\frac{(1-x)}{x}+\beta\frac{(-1-x)}{x}+\psi\frac{-2-x}{x}}P_2(\eta)\mathrm{d}\eta\\
\end{aligned}
\end{equation}
For $x\geq 0$ the probability $F_4^{xg4}(x)$ becomes:
\begin{equation}\label{eq:equation_160}
\begin{aligned}
    F_4^{xg4}(x)=&\int_{-\infty}^{\infty}\,\mathrm{d}\psi\,P_5(\psi)\int_{-\infty}^\infty\,\mathrm{d}\beta\,P_3(\beta)
    \int_{-\infty}^{\beta+2\psi}\,\mathrm{d}\xi\,P_1(\xi)
    \int_{-\xi-\beta-\psi}^{\infty}P_2(\eta)\mathrm{d}\eta+\\
    &\int_{-\infty}^{\infty}\,\mathrm{d}\psi\,P_5(\psi)\int_{-\infty}^\infty\,\mathrm{d}\beta\,P_3(\beta)
    \int_{\beta+2\psi}^{\infty}\,\mathrm{d}\xi\,P_1(\xi)
    \int_{\xi\frac{(1-x)}{x}+\beta\frac{(-1-x)}{x}+\psi\frac{-2-x}{x}}^{+\infty}P_2(\eta)\mathrm{d}\eta+\\
    &\int_{-\infty}^{\infty}\,\mathrm{d}\psi\,P_5(\psi)\int_{-\infty}^\infty\,\mathrm{d}\beta\,P_3(\beta)
    \int_{-\infty}^{\beta+2\psi}\,\mathrm{d}\xi\,P_1(\xi)
    \int_{-\infty}^{\xi\frac{(1-x)}{x}+\beta\frac{(-1-x)}{x}+\psi\frac{-2-x}{x}}P_2(\eta)\mathrm{d}\eta\\
    &\int_{-\infty}^{\infty}\,\mathrm{d}\psi\,P_5(\psi)\int_{-\infty}^\infty\,\mathrm{d}\beta\,P_3(\beta)
    \int_{\beta+2\psi}^{+\infty}\,\mathrm{d}\xi\,P_1(\xi)
    \int_{-\infty}^{-\xi-\beta-\psi}P_2(\eta)\mathrm{d}\eta\,.
\end{aligned}
\end{equation}
As above, it is easy to verify that $\lim_{x\rightarrow-\infty}F_3^{xg4}(x)=0$
and $\lim_{x\rightarrow+\infty}F_4^{xg4}(x)=1$.
The derivative of eq.~\ref{eq:equation_160} respect to $x$
gives the PDF of $x$:
\begin{equation}\label{eq:equation_170}
\begin{aligned}
    &P_{xg4}^l(x)=\\
    &\frac{1}{x^2}\Big[\int_{-\infty}^{\infty}\mathrm{d}\psi P_5(\psi)
    \int_{-\infty}^{+\infty}\mathrm{d}\beta P_3(\beta)\int_{\beta+2\psi}^{+\infty}\mathrm{d}\xi
    P_1(\xi)P_2(\xi\frac{1-x}{x}+\beta\frac{-1-x}{x}+\psi\frac{-2-x}{x})(-\beta+\xi-2\psi)+\\
    &\int_{-\infty}^{\infty}\,\mathrm{d}\psi\,P_5(\psi)
    \int_{-\infty}^{+\infty}\mathrm{d}\beta\,P_3(\beta)\int_{-\infty}^{\beta+2\psi}\mathrm{d}\xi\,
    P_1(\xi)P_2(\xi\frac{1-x}{x}+
    \beta\frac{-1-x}{x}+\psi\frac{-2-x}{x})(\beta-\xi+2\psi)\Big]
\end{aligned}
\end{equation}
With the transformation $\delta=\xi-\beta-2\psi$, eq.~\ref{eq:equation_170} becomes:
\begin{equation}\label{eq:equation_180}
\begin{aligned}
    P_{xg4}^l(x)=&\frac{1}{x^2}\Big[\int_{-\infty}^{\infty}\!\mathrm{d}\psi P_5(\psi)\!
    \int_{-\infty}^{+\infty}\!\mathrm{d}\beta P_3(\beta)\!\int_0^{+\infty}\!\mathrm{d}\delta
    P_1(\delta+\beta+2\psi)P_2(\delta\frac{1-x}{x}-2\beta-3\psi)\delta-\\
    &\int_{-\infty}^{\infty}\!\mathrm{d}\psi P_5(\psi)
    \int_{-\infty}^{+\infty}\!\mathrm{d}\beta P_3(\beta)\int_{-\infty}^{0}\!\mathrm{d}\delta P_1(\delta+\beta+2\psi)P_2(\delta\frac{1-x}{x}
    -2\beta-3\psi)\delta\Big]
\end{aligned}
\end{equation}
or better:
\begin{equation}\label{eq:equation_190}
    P_{xg4}^l(x)=\frac{1}{x^2}\Big[\!\int_{-\infty}^{\infty}\!\mathrm{d}\psi P_5(\psi)
    \int_{-\infty}^{+\infty}\!\mathrm{d}\beta P_3(\beta)\!\int_{-\infty}^{+\infty}\!\mathrm{d}\delta
    P_1(\delta+\beta+2\psi)P_2(\delta\frac{1-x}{x}-2\beta-3\psi)\mathrm{abs}(\delta)\!\Big]
\end{equation}
This can be recast in a form more appropriate for
the use with Gaussian PDFs of the signals:
\begin{equation}\label{eq:equation_190a}
    P_{xg4}^l(x)=\frac{1}{x^2}\Big[
    \int_{-\infty}^{+\infty}\mathrm{d}\delta\,\mathrm{abs}(\delta)\,
    \int_{-\infty}^{\infty}\,\mathrm{d}\psi\,P_5(\psi)\int_{-\infty}^{+\infty}\mathrm{d}\beta\,P_3(\beta)
    P_1(\delta+\beta+2\psi)P_2(\delta\frac{1-x}{x}-2\beta-3\psi)\Big]
\end{equation}

\subsection{Complete probability distribution for $P_{xg4}(x)$}
The two probability density functions, calculated up to now, are only a
part of the full probability distribution for the COG$_4$ algorithm.
In fact, the COG$_4$ algorithm is composed by the three
internal strips and the strip with maximum signal selected from the
two lateral strips ($<5>$ and $<4>$). We have to add this further condition.
When the left strip has a signal greater than the
right one  $P_{xg4}(x)$ converges toward  $P_{xg4}^l(x)$.
If the greater strip signal is the right one, $P_{xg4}(x)$
converges toward $P_{xg4}^r(x)$.

\begin{equation}\label{eq:equation_200}
\begin{aligned}
    &P_{xg4}(x)=\frac{1}{x^2}\Big[\\
    &\int_{-\infty}^\infty\mathrm{d}\psi P_5(\psi)\int_{\psi}^{\infty}\!\mathrm{d}\gamma P_4(\gamma)
    \int_{-\infty}^{+\infty}\mathrm{d}\beta P_3(\beta)\int_{-\infty}^{+\infty}\mathrm{d}\delta
    P_1(\delta+\beta-2\gamma)P_2(\delta\frac{1-x}{x}-2\beta+\gamma)\mathrm{abs}(\delta)+\\
    &\int_{-\infty}^\infty\mathrm{d}\gamma P_4(\gamma)\int_{\gamma}^{\infty}\!\mathrm{d}\psi P_5(\psi)
    \int_{-\infty}^{+\infty}\!\mathrm{d}\beta P_3(\beta)\!\int_{-\infty}^{+\infty}\!\mathrm{d}\delta
    P_1(\delta+\beta+2\psi)P_2(\delta\frac{1-x}{x}-2\beta-3\psi)\mathrm{abs}(\delta)\!\Big]\\
\end{aligned}
\end{equation}

Another form is given by the Fubini's theorem applied to the first two integrals:

\begin{equation}\label{eq:equation_200a}
\begin{aligned}
    &P_{xg4}(x)=\frac{1}{x^2}\Big[\\
    &\int_{-\infty}^\infty\mathrm{d}\gamma P_4(\gamma)\int_{-\infty}^{\gamma}\!\mathrm{d}\psi P_5(\psi)
    \int_{-\infty}^{+\infty}\mathrm{d}\beta P_3(\beta)\int_{-\infty}^{+\infty}\mathrm{d}\delta
    P_1(\delta+\beta-2\gamma)P_2(\delta\frac{1-x}{x}-2\beta+\gamma)\mathrm{abs}(\delta)+\\
    &\int_{-\infty}^\infty\mathrm{d}\psi P_5(\psi)\int_{-\infty}^{\psi}\!\mathrm{d}\gamma P_4(\gamma)
    \int_{-\infty}^{+\infty}\!\mathrm{d}\beta P_3(\beta)\!\int_{-\infty}^{+\infty}\!\mathrm{d}\delta
    P_1(\delta+\beta+2\psi)P_2(\delta\frac{1-x}{x}-2\beta-3\psi)\mathrm{abs}(\delta)\!\Big]\\
\end{aligned}
\end{equation}

With simple transformations (as $z=\delta/x$), the PDFs of
ref.~\cite{landi12} are easily recovered.
This last PDF and that with five strips of ref.~\cite{landi12} were
principally conceived for the extraction of the detector parameters
from the corresponding COG histograms with a self-consistent process.
The explicit PDFs for Gaussian additive noise require extensive use
of MATHEMATICA~\cite{MATHEMATICA} and are reported in ref.~\cite{landi12}.

\section{Summary and conclusions}

The probability density functions for center of gravity
algorithms with three strips and four strips are computed with
the classical method of differentiation of the corresponding cumulative
distributions. These derivations, through the cumulative distributions,
are consistent verifications of our published shorter methods.
The probability density function for the three strip algorithm
was extensively used in previous works for reproducing the
corresponding histograms of the center of gravity and testing
the detector parameters extracted from the data. In fact, the
simulations performed at orthogonal incidence do not
require these more complex probabilities, the two strips
formalism suffices. Almost always, the third and forth strips
have no signal information, but essentially noise.
Instead, the four and five strip algorithms can be used to a finer
tuning of the detector parameters extracted from the data,
the data histograms depend significantly on their values.
These non-linear dependencies allow the construction of a
self-consistent process for removing small artifacts of the
first order reconstructions.

%



\begin{thebibliography}{99}

\bibitem{landi08} Landi G.; Landi G. E. {\em The Cramer-Rao inequality to improve  of
                 the resolution of the least-squares method in track fitting}
                  {INSTRUMENTS} {\bf 2020}, 4(1), 2. https://doi.org/10.3390/instruments4010002

\bibitem{landi09} Landi G.; Landi G. E. {\em Generalized inequelities to optimizing the fitting
                  method for track reconstructions}, Physics {\bf 2020} 2(4)
                  https://doi.org/10.3390/physics2040035


\bibitem{landi10} Landi G.; Landi G. E.; {\em Probability Distributions of Positioning Errors
                   for Some Forms of Center-of-Gravity Algorithms.}
                   {\tt arXiv:2004.08975 [physics.ins-det]}


\bibitem{landi12} Landi G.; Landi G. E.; {\em Probability Distributions of Positioning Errors
                   for Some Forms of Center-of-Gravity Algorithms. Part II}
                   {\tt arXiv:2011.14474 [physics.ins-det]}

\bibitem{landi06} Landi, G.;  Landi G. E.   {\em Optimizing momentum resolution with a new fitting
                  method for silicon-strip detectors}  { INSTRUMENTS} {\bf 2018}, 2(4), 22
                  https://doi.org/10.3390/instruments2040022



\bibitem{gnedenko} B. V. Gnedenko "The Theory of Probability and Elements of Statistics"
                   (AMS Chelsea Publishing -Providence Rhode Island )


\bibitem{landi11} Landi G.; Landi G. E.; {\em Positioning Error Probability for Some Forms of
                 Center-of-Gravity Algorithms Calculated with the Cumulative Distributions. Part I.}
                  {\tt arXiv:2006.02934[physics.ins-det] }

\bibitem{landi05} Landi G.;  Landi G. E. {\em Improvement of track reconstruction with well tuned
                   robability distributions}  {\em JINST} 9 2014 P10006. {\tt arXiv:1404.1968[physics.ins-det]}
                   https://arxiv.org/abs/1404.1968


\bibitem{landi07} Landi G.; Landi G. E. {\em Beyond the $\sqrt{N}$-limit of the least
 squares resolution and the lucky-model} {\tt arXiv:1808.06708[physics.ins-det]}
 https://arxiv.org/abs/1808.06708.

\bibitem{matlab} MatLab 8 The MathWork Inc. Natic, MA, USA

\bibitem{rudi} Frühwirth R.; {\em Regression with Gaussian mixture models applied
to track fitting} {INSTRUMENTS } {\bf 2020}, 4(3) 25.

\bibitem{bernard} Bernard D. "Heteroscedasticity and angle resolution in
high-enery particle tracking: rivisiting "Beyond the $\sqrt{N}$ limits of
the least squares resolution and the lucky model", by G. Landi and G. E. Landi"
{\tt arXiv:2010.03451[physics.ins-det]}


\bibitem{landi01} G. Landi, {\em The center of gravity as an algorithm for position measurements}
Nucl. Instr. and Meth. {\bf A 485} (2002) 698 {\tt arXiv:1908.04447
[physics.ins-det]} https://arxiv.org/abs/1910.04447.


\bibitem{landi03} G. Landi, {\em Problems of position reconstruction in silicon
microstrip detectors} Nucl. Instr. and Meth. {\bf A 554} (2005) 226.



%
\bibitem{MATHEMATICA} MATHEMATICA 6 Wolfram Inc. Champaign IL, USA





\end{thebibliography}
\end{document}